# Sub-5 nm Gate-All-Around InP Nanowire Transistors Towards High-Performance Devices

Linqiang Xu,[1,2] Lianqiang Xu,[3] Qiuhui Li,[1] Shibo Fang,[1] Ying Li,[1] Ying Guo,[4] Aili Wang,[5,6] Ruge Quhe,[7*] Yee Sin Ang,[2*] and Jing Lu[1,8,9,10,11*]

[1]State Key Laboratory of Mesoscopic Physics and Department of Physics, Peking University, Beijing 100871, P. R. China

[2]Science, Mathematics and Technology, Singapore University of Technology and Design (SUTD), 8 Somapah Road, Singapore 487372, Singapore

[3]School of Physics and Electronic Information Engineering, Engineering Research Center of Nanostructure and Functional Materials, Ningxia Normal University, Guyuan 756000, P. R. China

[4]School of Physics and Telecommunication Engineering, Shaanxi Key Laboratory of Catalysis, Shaanxi University of Technology, Hanzhong, 723001, People's Republic of China

[5]College of Information Science and Electronic Engineering, Zhejiang University, Hangzhou, China

[6]Zhejiang University - University of Illinois at Urbana-Champaign Institute, Zhejiang University, Haining, China

[7]State Key Laboratory of Information Photonics and Optical Communications and School of Science, Beijing University of Posts and Telecommunications, Beijing 100876, P. R. China

[8]Collaborative Innovation Center of Quantum Matter, Beijing 100871, P. R. China

[9]Beijing Key Laboratory for Magnetoelectric Materials and Devices, Beijing 100871, P. R. China

[10]Peking University Yangtze Delta Institute of Optoelectronics, Nantong 226000, P. R. China

[11]Key Laboratory for the Physics and Chemistry of Nanodevices, Peking University, Beijing 100871, P. R. China

*Corresponding Authors: quheruge@bupt.edu.cn, yeesin_ang@sutd.edu.sg, jinglu@pku.edu.cn

## Abstract

Gate-all-around (GAA) nanowire (NW) field-effect transistor (FET) is a promising device architecture due to its superior gate controllability than that of the conventional FinFET architecture. The significantly higher electron mobility of indium phosphide (InP) NW than silicon NW makes it particularly well-suited for high-performance (HP) electronics applications. In this work, we perform an *ab initio* quantum transport simulation to investigate the performance limit of sub-5-nm gate length ($L_g$) GAA InP NW FETs. The GAA InP NW FETs with $L_g$ of 4 nm can meet the International Technology Roadmap for Semiconductors (ITRS) requirements for HP devices from the perspective of on-state current, delay time, and power dissipation. We also investigate the impact of strain on 3-nm-$L_g$ GAA InP NW FETs. The application of tensile strain results in a remarkable increase of over 60% in the on-state current. These results highlight the potential of GAA InP NW FETs for HP applications in the sub-5-nm $L_g$ region.

# 1. Introduction

Fin field-effect transistor (FinFET) is the dominant device architecture in current mainstream silicon (Si)-based technology.[1-3] However, Si FinFET is reaching its limit at the 5 nm technology node (18 nm gate length) due to the severe short-channel effect (SCE).[4-6] Gate-all-around (GAA) FET is considered a replacement for FinFET owing to its superior gate controllability.[1, 4, 6, 7] The better gate control can be quantitatively described by the natural length $\lambda$ ($\lambda=\sqrt{\alpha t_{ch} t_{ox} \varepsilon_{ch}/\varepsilon_{ox}}$), which represents the penetrating depth of source/drain electric field inside the channel.[8, 9] Here, $\alpha$ is the structure parameter (1/3 for FinFET and 1/4 for GAA FET), and $t_{ch}/\varepsilon_{ch}$ ($t_{ox}/\varepsilon_{ox}$) represents the thickness/permittivity of channel (dielectric oxide). GAA FET exhibits smaller $\lambda$, indicating better gate control. On the other hand, the one-dimensional (1D) channel in the GAA structure is more conducive to achieving ballistic transport as compared to the two-dimensional (2D) and bulk channels because backscattering only occurs in one direction.[10] As a result, the 1D channel generally possesses higher carrier mobility than other 2D and 3D semiconductors. For instance, the carrier mobility of the 1D carbon nanotube is up to 100000 cm$^2$ V$^{-1}$ s$^{-1}$, significantly surpassing that of bulk Si (1450 cm$^2$ V$^{-1}$ s$^{-1}$) and 2D semiconductors (usually less than 10000 cm$^2$ V$^{-1}$ s$^{-1}$).[11-13]

Extensive research has been conducted on 1D Si nanowire (NW) GAA FETs.[14-16] However, the relatively lower electron mobility ($\mu_e$) of Si NW (about 28-770 cm$^2$ V$^{-1}$ s$^{-1}$) has limited its application for high-performance (HP) electronics.[17, 18] III-V materials are considered promising substitutions of Si in accordance with the International Technology Roadmap for Semiconductors (ITRS) and International Roadmap for Devices and Systems (IRDS) due to their higher $\mu_e$ and injection velocity than Si.[6, 19, 20] Among the III-V materials, indium phosphide (InP) NW exhibits $\mu_e$ of up to 2000 cm$^2$ V$^{-1}$ s$^{-1}$, making it very potential for achieving HP goals outlined by ITRS and IRDS.[21] Experimentally, the InP NW FETs with gate lengths ($L_g$) larger than 300 nm have been successfully fabricated.[22-26] For example, Ho et al. showed that the 2-μm-$L_g$ InP NW FET exhibits a current on-off ratio of up to 10$^6$.[25] To further shorten $L_g$, it is crucial to reduce the NW diameter ($D_{NW}$) for suppressing SCE. This is because reducing $D_{NW}$ helps decrease $\varepsilon_{ch}$ (consequently $\lambda$) and thus enhances the gate controllability, as demonstrated in the Si NW system.[27] However, to our best knowledge, the smallest $D_{NW}$ achieved in fabricated InP NW FETs thus far exceeds 10 nm.[21-24] The performance characteristics and limits of InP NW FETs with ultra-short $L_g$ and ultra-small $D_{NW}$ remain an open question thus far.

In this work, we study the device properties of sub-5 nm $L_g$ GAA InP NW FET with $D_{NW}$ of 1.6 nm based on the *ab initio* quantum transport simulation. Our findings show that the on-state current of 4-nm-$L_g$ GAA InP NW FET can reach the ITRS HP targets. Moreover, the delay time and power dissipation of GAA InP NW FET satisfy the ITRS HP demands as $L_g$ is scaled down to 3 and 1 nm, respectively. Inspired by the positive effect of strain on the III-V NW device performance, we apply uniaxial strain on the 3-nm-$L_g$ GAA InP NW FET. Intriguingly, the application of tensile strain leads to an improvement of over 60% in the on-state current due to the strain-induced reduction of the electron effective mass in InP NW. Our findings reveal GAA ultra-small InP NW FET as a compelling building block for next-generation HP electronics applications.

## 2. Method

The transport properties of InP NW FETs are simulated based on the density functional theory coupled with the non-equilibrium Green's function (NEGF) method, implemented in the Quantum ATK 2020 package.[28-32] For a FET, the device contains the electrode and central regions. The electrode region is semi-finite and in the equilibrium state, while the central region is in a non-equilibrium state when the bias voltage ($V_b$) is applied. The interaction between these two regions can be described by the self-energy $\Sigma_{k_{//}}^{l/r}$, where $k_{//}$ is a surface-parallel reciprocal lattice vector, and $l/r$ represents the left/right electrodes. $\Sigma_{k_{//}}^{l/r}$ can be calculated via $\Sigma_{k_{//}}^{l/r} = \tau_{k_{//}}^{l/r} G_{k_{//}}^{l/r}(E) (\tau_{k_{//}}^{l/r})^+$, in which $G_{k_{//}}^{l/r}(E)$ is the retarded Green's function of the left/right electrode and $\tau_{k_{//}}^{l/r}$ is the coupling Hamiltonian between the electrode and central regions.

Based on the self-energy and central region Hamiltonian $H_{k_{//}}$, we can obtain the central region retarded Green's function $G_{k_{//}}(E)$ by the formula $G_{k_{//}}(E) = [(E + i0^+)I - H_{k_{//}} - \Sigma_{k_{//}}^{l} - \Sigma_{k_{//}}^{r}]^{-1}$, where $0^+$ and $I$ are the positive infinitesimal and unit matrix, respectively. The advanced central Green's function is given by $G_{k_{//}}^{\dagger}(E) = [G_{k_{//}}(E)]^{\dagger}$. Additionally, the broadening matrix $\Gamma_{k_{//}}^{l/r}(E)$ is the imaginary part of $\Sigma_{k_{//}}^{l/r}$ and can be obtained as $\Gamma_{k_{//}}^{l/r}(E) = i[\Sigma_{k_{//}}^{l/r} - (\Sigma_{k_{//}}^{l/r})^{\dagger}]$. Thus, the $k$-dependent transmission coefficients can be calculated by the following equation:

$$T_{k_{//}}(E) = Tr[\Gamma_{k_{//}}^{l}(E) G_{k_{//}}(E) \Gamma_{k_{//}}^{r}(E) G_{k_{//}}^{\dagger}(E)] \quad (1)$$

By averaging the transmission coefficients over the irreducible Brillouin zone, we can get the transmission function $T(E)$. Subsequently, the drain current ($I_{ds}$) is obtained according to the Landauer–Büttiker formula:

$$I_{ds} = \frac{2e}{h} \int_{-\infty}^{+\infty} \left[ f_D(E - \mu_D) - f_S(E - \mu_S) \right] T(E) \, dE \tag{2}$$

in which $f_S$ ($f_D$) stands for the source (drain) Fermi-Dirac distribution functions, and $\mu_S$ ($\mu_D$) denotes the source (drain) electrochemical potential. The ballistic transport is adopted in our calculation since it dominates over scattering in the sub-10 nm $L_g$ region.[33, 34] The PseudoDojo pseudopotential, temperature of 300 K, and $k$-point meshes of 1×1×137 are set. We apply the Neumann, Neumann, and Dirichlet boundary conditions for the $x$, $y$, and $z$ directions (Figure 1(c)), respectively. Throughout the simulation, the generalized gradient approximation (GGA) in the Perdew-Burke-Ernzerhof (PBE) form is employed.[35]

The DFT-GGA method provides accurate bandgap evaluation in FETs due to the screening effect from the dielectric layer and doping carrier.[9, 36, 37] On the one hand, the electron-electron interaction is shielded by the dielectric layer in the device environment. Previous studies have demonstrated that the renormalized GW bandgap of the monolayer (ML) $MoS_2$, sandwiched between two $HfO_2$ dielectric layers, is 1.9 eV,[36] which is comparable to the DFT-GGA bandgap of 1.76 eV.[38] On the other hand, the injected carriers from the electrodes also contributed to the screening of the electron-electron interaction. For example, the calculated bandgap of the degenerately doped ML $MoSe_2$ at the DFT-GGA level is 1.52 eV, which is in agreement with the GW approximation (1.59 eV) and experimental results (1.58 eV).[39-41] Besides, the simulated device performance using the DFT-NEGF method exhibits great consistency with the experimental results. For instance, the simulated performance of the 5-nm-$L_g$ single-walled carbon nanotube (CNT), including the transfer curves, on-state current, delay time, and power dissipation, agrees well with experimental data.[42] DFT-NEGF simulations hence provide a powerful predictive tool to investigate the performance and optimal design of ultrascaled nanochannel transistors.

## 3. Results

### 3.1 Electronic and Device Structure

Bulk InP exhibits two kinds of phases including the wurtzite (WZ) and zinc-blend (ZB) phases. Previous study showed that InP NW prefers the WZ phase when $D_{NW}$ is less than 7 nm.[43] Thus, we use the bulk WZ phase InP to construct the InP NW. The optimized structure is shown in Figure 1(a). To eliminate the dangling bonds, hydrogen atoms are employed to passivate the edge of InP NW. The optimized lattice parameter along the transport ($z$) direction is $c$ = 6.90 Å. We also calculate the band structure of the InP NW by Quantum ATK, as depicted in Figure 1(b). The InP NW exhibits a direct band gap of 2.21 eV. The effective mass ($m^*$) can be extracted from the band structure, and the obtained electron $m^*$ of InP NW along the transport direction is 0.227 $m_0$, in which $m_0$ is the mass of an electron.

Figure 1(c) illustrates the device structure of GAA InP NW FET. The source and drain electrodes consist of highly doped InP NW, while the channel is modeled by the intrinsic InP NW. Incorporating symmetrically distributed underlap (UL) structures (segments of the channel) can enhance device performance by decreasing the impact of the source/drain on the channel and reducing leakage current. However, an excessively long UL length can deteriorate the gate controllability. Hence, an optimal UL length is necessary to achieve superior device performance. The gate dielectric uses the $SiO_2$ with a thickness of 0.41 nm based on the ITRS 2013 version standard. We use the ITRS 2013 version, rather than the latest IRDS 2022 version, as the benchmark because ITRS 2013 possesses shorter $L_g$ scaling (5 nm for ITRS 2013 and 12 nm for IRDS 2022) and stricter requirements for the device properties. Hereafter, "ITRS 2013 version" and "IRDS 2022 version" are denoted as "ITRS" and "IRDS", respectively.

The diameter and perimeter normalizations in the cross-section of the nanowire are widely used for the current normalization. When $D_{NW}$ is less than 10 nm, the current is typically normalized by diameter due to the concentrated electrical conduction in the core of the nanowire.[44, 45] This can be explained by a simple model of a particle confined in an infinite cylindrical potential well, as shown in the Si NW FETs system.[27] For the 1.6-nm-$D_{NW}$ InP NW FET, we calculate its current density with $L_g$ of 3 nm (Figure 1(d)). The current is primarily concentrated in the core of the ultranarrow InP NW FET. Therefore, unless explicitly stated otherwise, we adopt diameter normalization in our subsequent simulations.

## 3.2 On-state Current

The on-state current ($I_{on}$) is a crucial indicator for assessing the device performance. It can be determined from the transfer curves at the on-state voltage ($V_g^{on}$) point. $V_g^{on}$ is calculated by $V_g^{on} = V_g^{off} + V_{dd}$, where $V_g^{off}$ and $V_{dd}$ are the off-state and supply voltages, respectively. According to the ITRS HP standard, $V_{dd}$ and off-state current ($I_{off}$) are 0.64 V and 0.1 μA/μm, respectively. $V_g^{off}$ can be derived from the transfer characteristics at the $I_{off}$ point. The doping type and concentration of the electrodes significantly impact the performance of InP NW FETs. Experimentally, InP NW is predominantly employed as the channel material of *n*-type devices.[21, 22, 24, 25] Hence, we focus on electron doping in the subsequent simulation. We adopt the atomic compensation charge method for doping, and further details can be found in the supplementary material. Three levels of doping concentrations ($3.6 \times 10^{19}$, $7.2 \times 10^{19}$, and $3.6 \times 10^{20}$ cm$^{-3}$) are tested, as depicted in Figure S1. A doping concentration of $3.6 \times 10^{19}$ cm$^{-3}$ is chosen due to its significantly higher $I_{on}$ than the other two concentrations (2730/354/122 μA/μm for $3.6 \times 10^{19}/7.2 \times 10^{19}/3.6 \times 10^{20}$ cm$^{-3}$).

Based on the tested doping concentration, we calculated the transfer characteristics of InP NW FETs at $L_g$ = 5/4/3/1 nm, as displayed in Figure S2. The extracted $I_{on}$ for different $L_g$ and UL lengths is presented in Table 1. In Figure 2(a), we plot the UL-optimized $I_{on}$ against $L_g$ of the simulated GAA InP NW, GAA Si NW ($D_{NW}$ = 1.0 nm), GAA CNT ($D_{NW}$ = 0.6 nm), and double-gated (DG) ML InP FETs. [27, 42, 46] For the InP NW FETs, $I_{on}$ can achieve the ITRS HP target until $L_g$ is down to 4 nm. By contrast, the scaling limit for the Si NW FET, CNT, and ML InP FETs are 3, 2, and 2 nm, respectively. The superior scaling behavior of the Si NW, CNT, and ML InP than InP NW is mainly attributed to the smaller electron $m^*$. Previous studies have revealed the relationship between $I_{on}$ and $m^*$.[9, 37, 46-48] With increasing $m^*$, $I_{on}$ decreases first when $m^* < 0.4$ $m_0$ due to the reduction of electron velocity, while it increases when $m^* > 0.4$ $m_0$ because of the improved density of states (DOS). Taking $L_g$ = 3 nm as an example, a similar trend is observed in Figure 2(b). Thus, a smaller $m^*$ of Si NW (0.127 $m_0$), CNT (0.172 $m_0$), and ML InP (0.093 $m_0$) than InP NW (0.277 $m_0$) results in better $I_{on}$ performance.

### 3.3 Gate Controllability

The subthreshold swing (*SS*) and transconductance ($g_m$) measure the gate controllability in subthreshold and superthreshold regions, respectively. A smaller *SS* and larger $g_m$ indicate better gate control. *SS* can be calculated by the formula $SS = \frac{\partial V_g}{\partial \lg I}$, and $g_m$ is defined as $g_m = \frac{dI}{dV_g}$. Figures 3(a) and 3(b) display *SS* and $g_m$ against $L_g$ with different UL lengths, respectively. *SS* decreases with increasing $L_g$ at the same UL lengths, indicating the improvement of gate controllability. For example, *SS* reduces by about 35% ($L_g$ from 3 to 5 nm), 70% ($L_g$ from 1 to 5 nm), and 58% ($L_g$ from 1 to 5 nm) for UL lengths of 0, 1, and 2 nm, respectively. Two reasons can explain this variation trend. Firstly, increasing $L_g$ extends the total channel length, which mitigates the short-channel effect and thus enhances gate control. Secondly, enlarging $L_g$ improves the proportion of the channel under the gate when the UL length is fixed. The enhancement of gate controllability is further demonstrated by $g_m$. For UL lengths of 0/1/2 nm, $g_m$ is increased by 3.3 ($L_g$ from 3 to 5 nm)/425 ($L_g$ from 1 to 5 nm)/5.8 ($L_g$ from 1 to 5 nm) times.

To further illustrate the variation of gate controllability with $L_g$, we calculate the local DOS (LDOS) and spectrum current of the InP NW FETs at $L_g$ = 1 and 4 nm (UL length of 2 nm), as depicted in Figure 4. The electron barrier height ($\Phi$) is defined as the difference between the Fermi level of drain ($\mu_d$) and the maximal conduction band edge of channel. Based on the maximal conduction band edge, the total spectrum current *I* can be separated into two components: tunneling current ($I_{tunnel}$) and thermionic current ($I_{therm}$). For $L_g$ = 1 nm, the switch from off-state to on-state in the InP NW device reduces $\Phi$ by 0.33 eV (from 0.38 eV to 0.05 eV). By contrast, $\Phi$ is decreased by 0.38 eV (from 0.36 eV to -0.02 eV) in the off-on switching process at $L_g$ = 4 nm due to the improved gate controllability. During the off-on switching process, $I_{tunnel}$ reduces for both gate lengths, while $I_{therm}$ enhances. The InP NW FET with 4 nm $L_g$ exhibits a larger $I_{therm}$ at the on-state than 1 nm $L_g$ because of the smaller $\Phi$.

### 3.4 Delay Time, Power Dissipation, and Energy-Delay Product

In FET, the charge storage capacity in the channel is characterized by the total capacitance $C_t$. A smaller $C_t$ indicates smaller charge storage capacity and thus faster switching ability. $C_t$ comprises gate capacitance $C_g = \partial Q_{ch}/\partial V_g$ ($Q_{ch}$ is the total charge in the channel) and fringing capacitance $C_f$. In terms of the ITRS criterion, $C_f$ is twice $C_g$, and hence $C_t$ is three times $C_g$. The $C_t$ values of the InP NW FETs with different gate and UL lengths are given in Table 1. Figure 5(a) shows the UL-optimized $C_t$ versus $L_g$ of the InP NW, Si NW, CNT, and ML InP FETs.[27, 42, 46] For the GAA InP NW FETs, $C_t$ decreases with reducing $L_g$. The optimal $C_t$ at all $L_g$ (0.12-0.55 fF/μm) can satisfy the ITRS HP demand (0.6 fF/μm). The InP NW FETs possess higher $C_t$ than the other FETs at all gate lengths except for 1-nm-$L_g$ Si NW FET.

The delay time $\tau$ ($\tau = C_t V_{dd}/I_{on}$) describes the *on-off* switching speed in a transistor. $\tau$ is directly proportional to $C_t$ and $V_{dd}$ (fixed at 0.64 V) while inversely proportional to $I_{on}$. A smaller $\tau$ is preferred as it implies a faster switching speed. The UL-optimized $\tau$ of the InP NW, Si NW, CNT, and ML InP FETs at different $L_g$ is illustrated in Figure 5(b).[27, 42, 46] When $L_g$ is scaled down to 3 nm, $\tau$ of the GAA InP NW FETs meets the ITRS HP criterion. At $L_g$ = 1 nm, the relatively low $I_{on}$ (48 μA/μm) results in a relatively high $\tau$ (2.14 ps), which surpasses the ITRS criterion. $\tau$ of the GAA InP NW FETs exceeds the other three kinds of FETs at all $L_g$ except for 1-nm-$L_g$ Si NW FET due to the higher $C_t$ and smaller $I_{on}$.

Energy consumption, usually characterized by the power dissipation PDP = $V_{dd} I_{on} \tau = C_t V_{dd}^2$, is a critical factor in evaluating the performance of a transistor. A smaller PDP indicates lower energy costs. Figure 5(c) plots the UL-optimized PDP of the InP NW, Si NW, CNT, and ML InP FETs.[27, 42, 46] The variation trend of PDP is consistent with $C_t$ for all the FETs since it is in proportion to $C_t$. In terms of the ITRS HP standard (0.24 fJ/μm), The gate scaling limit of InP NW FETs is 1 nm (optimal PDP is 0.066-0.22 fJ/μm for $L_g$ = 1-5 nm). PDPs of the InP NW FETs are almost larger than the other counterparts owing to the higher $C_t$.

A logic device necessitates both faster switching speed (smaller $\tau$) and lower energy consumption (smaller PDP). Such characteristics can be captured by the energy-delay product EDP = $\tau \times$ PDP. EDPs of the sub-5-nm-$L_g$ InP NW, Si NW, CNT, and ML InP FETs are compared in Figure 5(d).[27, 42, 46] The dashed lines represent the ITRS and IRDS criteria for EDP on the 2028 horizon. All the InP NW FETs satisfy the IRDS and stricter ITRS demands. Since the InP NW FETs own higher $\tau$ and PDP than CNT and ML InP FETs at all $L_g$, their corresponding EDPs also exceed the other counterparts. Compared to the Si NW FETs, the InP NW FET possesses higher EDP at $L_g$ = 3 and 5 nm due to larger $\tau$ and PDP while it exhibits smaller EDP at $L_g$ = 1 nm because of significantly lower $\tau$.

### 3.5 Strain Effect

Strain engineering has attracted much attention for boosting the performance of Si-based devices.[49, 50] Previous experimental studies demonstrated the uniaxial strain could effectively improve the carrier mobility and device performance of both Si and III-V NWs.[51-54] Motivated by these positive effects, we apply the uniaxial strain including the compression (-2%) and tension (2%) to the InP NW, as displayed in Figure 6(a). The impact of strain effect on the electron effective mass $m^*$ is investigated first (Figure 6(b)). Under -2% compressive strain, $m^*$ increases from 0.277 $m_0$ to 0.298 $m_0$ (approximately 7.5%). In contrast, implementing 2% tensile strain leads to a decrease in $m^*$ to 0.211 $m_0$ (approximately 24%). All $m^*$ values within the -2% to 2% strain range are smaller than 0.4 $m_0$. As discussed earlier, $I_{on}$ increases with reducing $m^*$ within this range. Our results further confirm this trend by applying the uniaxial strain to the 3-nm-$L_g$ InP NW FETs, as presented in Figure 6(c). The -2% compression results in the decrement of $I_{on}$ by about 54%. Nevertheless, $I_{on}$ is raised by over 60% with 2% tension strain.

## 4. Conclusion

In conclusion, we conducted the investigation of sub-5 nm $L_g$ GAA InP NW FETs with ultra-small diameter for the first time by the *ab initio* quantum transport simulation. $I_{on}$ of the InP NW FETs meets the ITRS HP criteria until $L_g$ is reduced to 4 nm. Moreover, the scaling limit of $\tau$ and PDP is down to 3 and 1 nm in terms of ITRS HP demands, respectively. Encouraged by the positive impact of uniaxial strains on nanowire device performance, we apply the uniaxial compressive and tensile strain to the 3-nm-$L_g$ InP NW FETs. Remarkably, with a 2% tensile strain, $I_{on}$ is increased by over 60% due to the reduced electron effective mass. Thus, our studies reveal the great potential of InP NW for sub-5 nm $L_g$ HP electronics.

## Acknowledgment

This work was supported by the National Natural Science Foundation of China (No. 91964101, No. 12274002, and No. 12164036), the Ministry of Science and Technology of China (No. 2022YFA1200072, No.2022YFA1203904), the China Scholarship Council, the Fundamental Research Funds for the Central Universities, the Natural Science Foundation of Ningxia of China (No. 2020AAC03271), the youth talent training project of Ningxia of China (2016), and the High-performance Computing Platform of Peking University. This project is also supported by the Open Fund of IPOC (BUPT). Y. S. A. and L. X. (Linqiang Xu) acknowledge the supports of Singapore University of Technology and Design Kickstarter Initiatives (SKI) under the Award No. SKI 2021_01_12 and SUTD Startup Research Grant (SRG SCI 2021 163). Y. S. A. is also supported by the SUTD-ZJU IDEA Thematic Research Grant Exploratory Project (SUTD-ZJU (TR) 202203). The computational work for this article was partially performed on resources of the National Supercomputing Centre, Singapore (https://www.nscc.sg).

## Data Availability Statement

The data that support the findings of this study are available from the corresponding authors upon reasonable request.

## Conflict of Interest

The authors declare no conflict of interest.

**Table 1.** Benchmark of the device performance for the *n*-type GAA InP NW FETs with $D_{NW}$ of 1.6 nm against the ITRS HP standards.

|  | $L_g$ (nm) | $L_{UL}$ (nm) | $I_{on}$ (μA/μm) | SS (mV/dec) | $g_m$ (μS/μm) | $C_t$ (fF/μm) | τ (ps) | PDP (fJ/μm) |
|---|---|---|---|---|---|---|---|---|
| *n*-type HP | 5 | 0 | 2730 | 135 | 3025 | 0.96 | 0.22 | 0.39 |
|  |  | 1 | 1510 | 82 | 4694 | 0.77 | 0.33 | 0.32 |
|  |  | 2 | 1469 | 89 | 5043 | 0.55 | 0.24 | 0.22 |
|  | 4 | 0 | 225 | 174 | 1906 | 0.58 | 1.66 | 0.24 |
|  |  | 1 | 857 | 139 | 3078 | 0.55 | 0.41 | 0.23 |
|  |  | 2 | 984 | 102 | 3250 | 0.51 | 0.34 | 0.21 |
|  | 3 | 0 | 20 | 210 | 699 | 0.34 | 10.7 | 0.14 |
|  |  | 1 | 402 | 170 | 1253 | 0.40 | 0.64 | 0.16 |
|  |  | 2 | 804 | 125 | 2203 | 0.37 | 0.30 | 0.15 |
|  | 1 | 0 | / | / | / | / | / | / |
|  |  | 1 | 1.75 | 278 | 11 | 0.12 | 0.64 | 0.16 |
|  |  | 2 | 48 | 213 | 741 | 0.16 | 2.15 | 0.066 |
| ITRS | 5.1 | / | 900 | / | / | 0.6 | 0.423 | 0.24 |

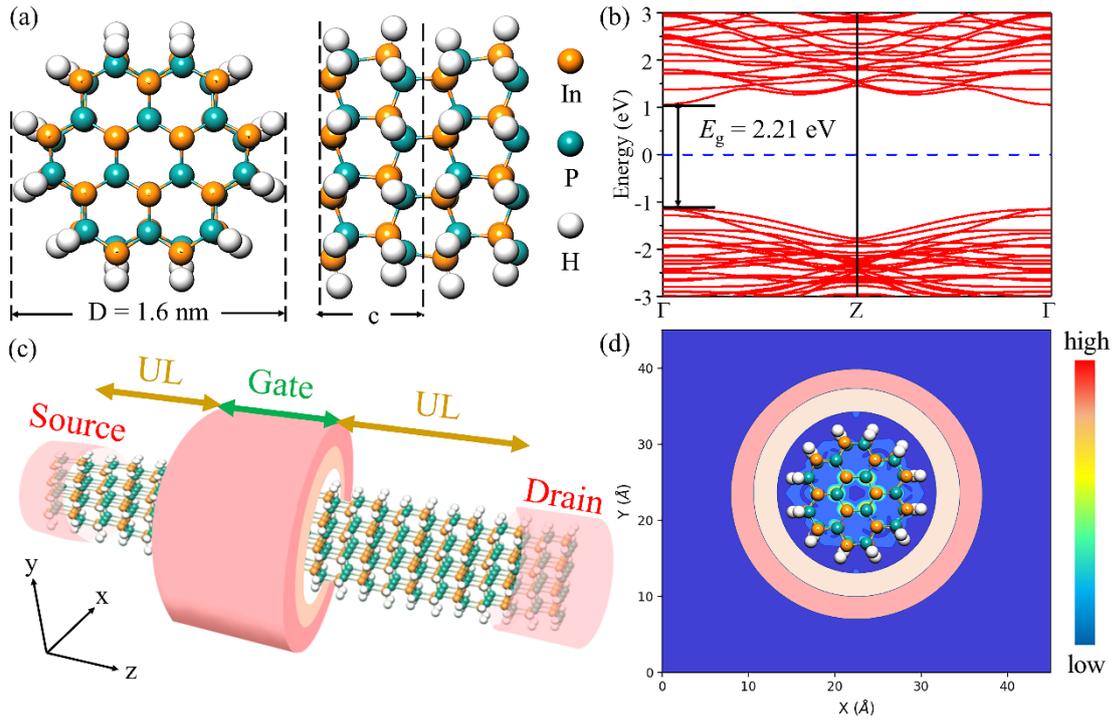

**Figure 1.** (a) Top view and side view of the InP NW with $D_{NW}$ of 1.6 nm. (b) Calculated band structure of the InP NW. (c) Schematic view of the GAA InP NW FETs. (d) The current density of GAA InP NW FETs on the X-Y plane.

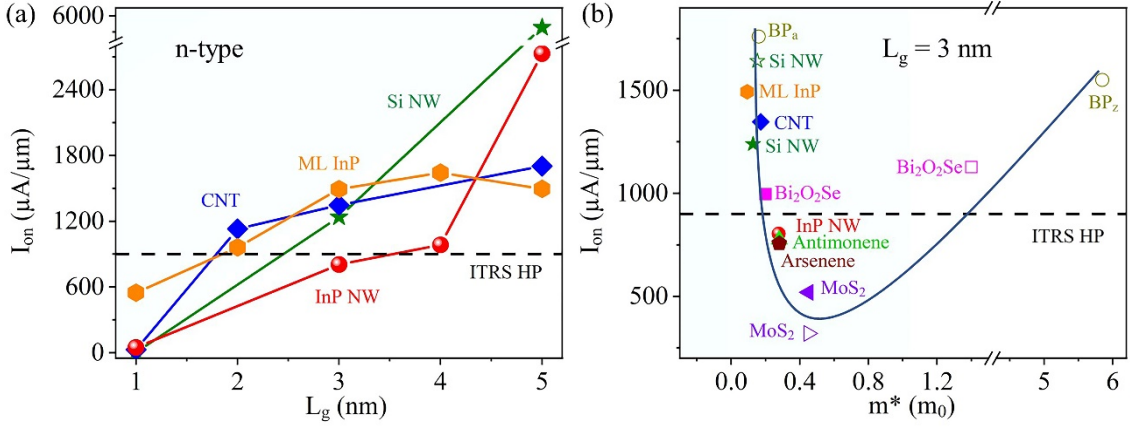

**Figure 2.** (a) Gate length scaling of UL-optimized $I_{on}$ for the *n*-type simulated GAA InP NW FETs (red). For comparison, the optimal values of the simulated GAA Si NW ($D_{NW}$ = 1.0 nm), GAA CNT FET ($D_{NW}$ = 0.6 nm), and DG ML InP FETs are also displayed.[27, 42, 46] (b) HP $I_{on}$ of the low-dimensional material FETs at $L_g$ = 3 nm against $m^*$ of the low-dimensional material.[27, 42, 46, 55-58] The blue lines are guides to the eyes. The solid and hollow symbols represent the *n*-type and *p*-type FETs, respectively.

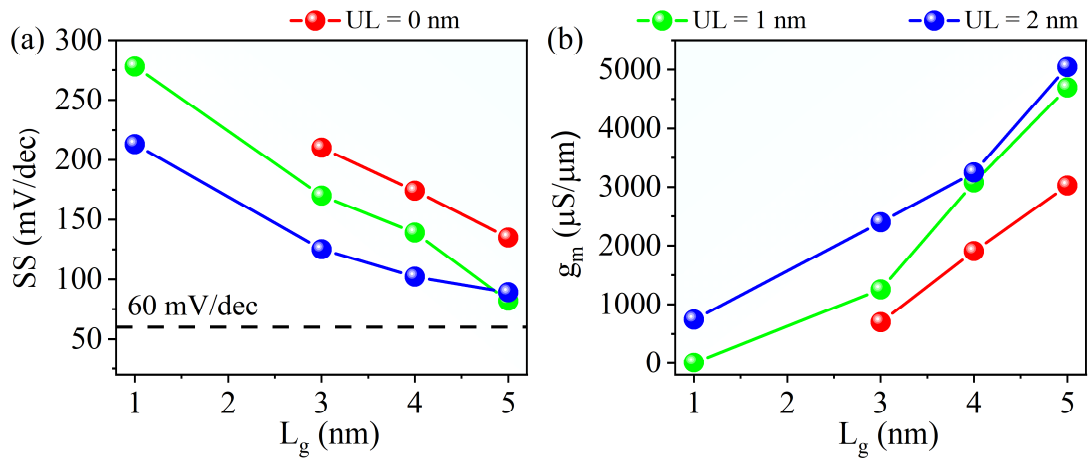

**Figure 3.** (a) Subthreshold swing SS and (b) transconductance $g_m$ of the GAA InP NW FETs as a function of $L_g$. The black dashed line represents the theoretical limit of SS at room temperature.

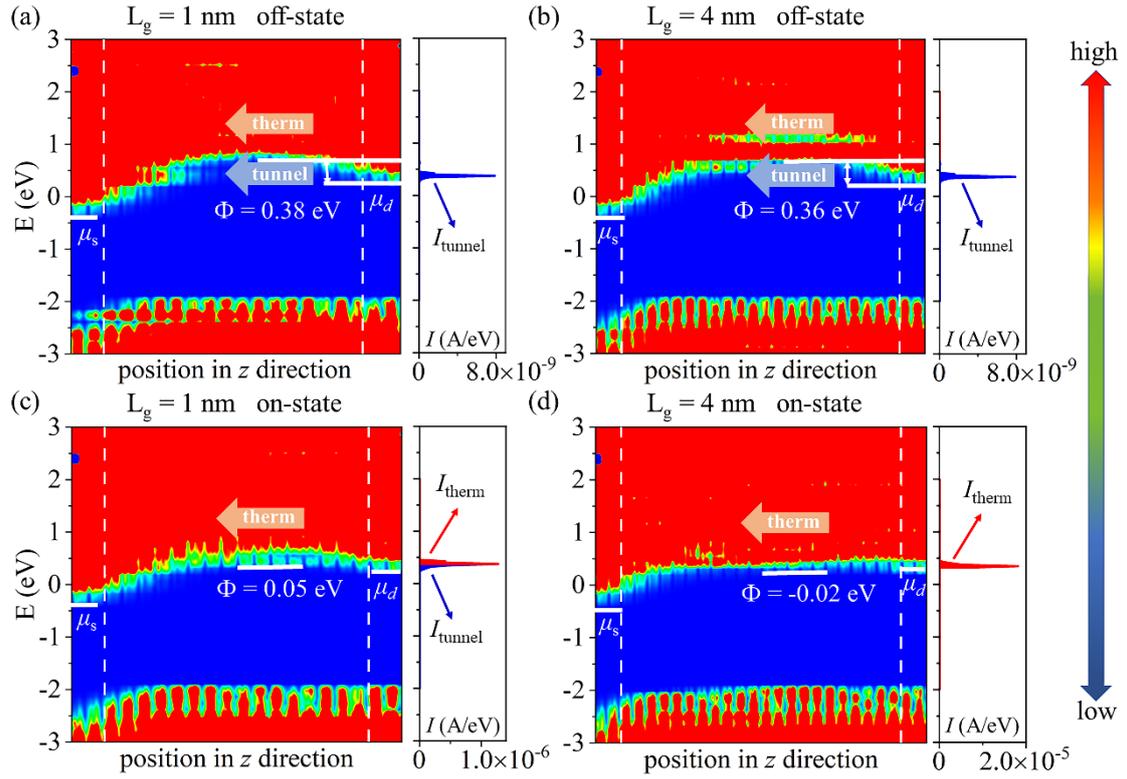

**Figure 4.** Local density of states (LDOS) and spectral current of the GAA InP NW FETs (UL length of 2 nm) at the (a) off-state of $L_g$ = 1 nm, (b) off-state of $L_g$ = 4 nm, (c) on-state of $L_g$ = 1 nm, and (d) on-state of $L_g$ = 4 nm. $\mu_s$, $\mu_d$, and $\Phi$ depict the Fermi level of source, Fermi level of drain, and electron barrier height. The white dashed lines show the boundary between source/drain and channel.

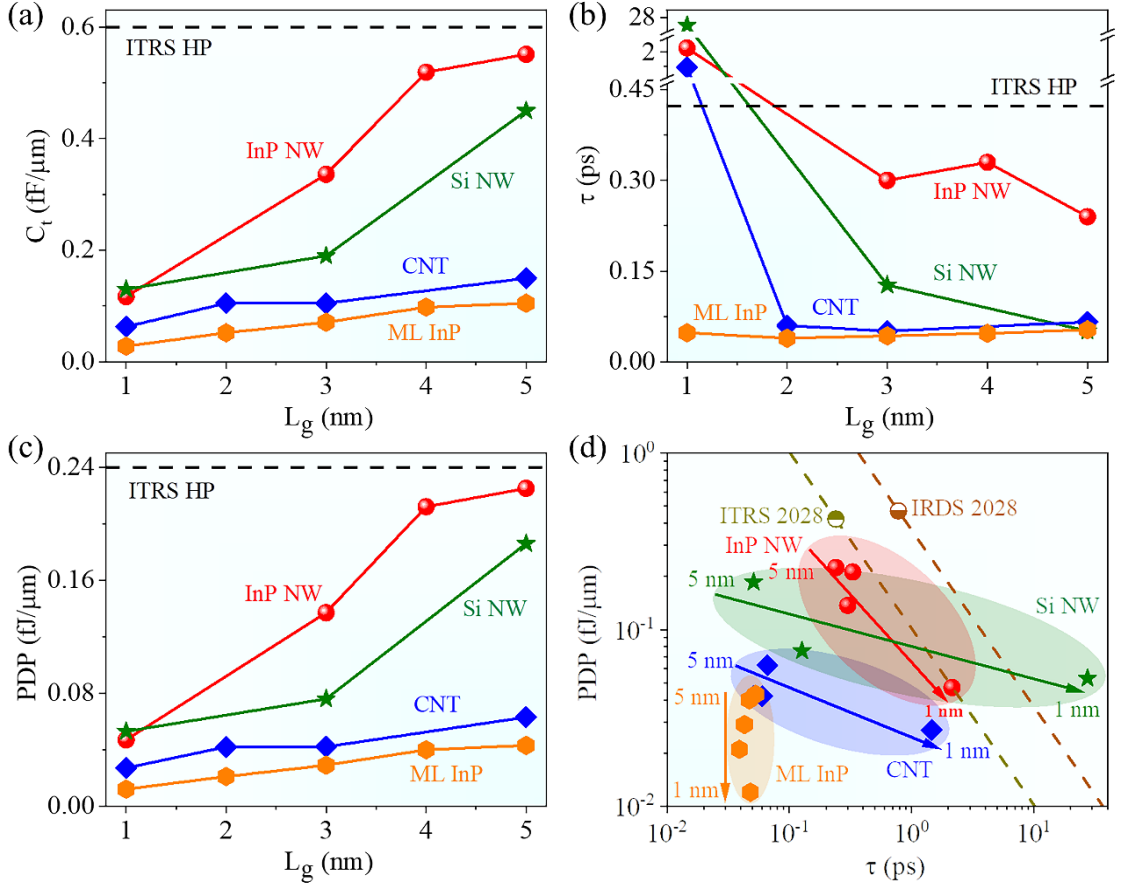

**Figure 5.** UL-optimized (a) total capacitance $C_t$, (b) delay time $\tau$, and (c) power dissipation PDP of the simulated GAA InP NW FETs against the gate length. For comparison, the optimal values of the simulated Si NW, CNT, and ML InP FETs are also shown. (d) PDP versus $\tau$ in the InP NW FETs against those of Si, CNT, and ML InP FETs with $L_g$ from 5 to 1 nm for the HP applications. The dashed lines indicate the ITRS and IRDS requirements for the energy-delay product EDP = $\tau \times$ PDP.[27, 42, 46]

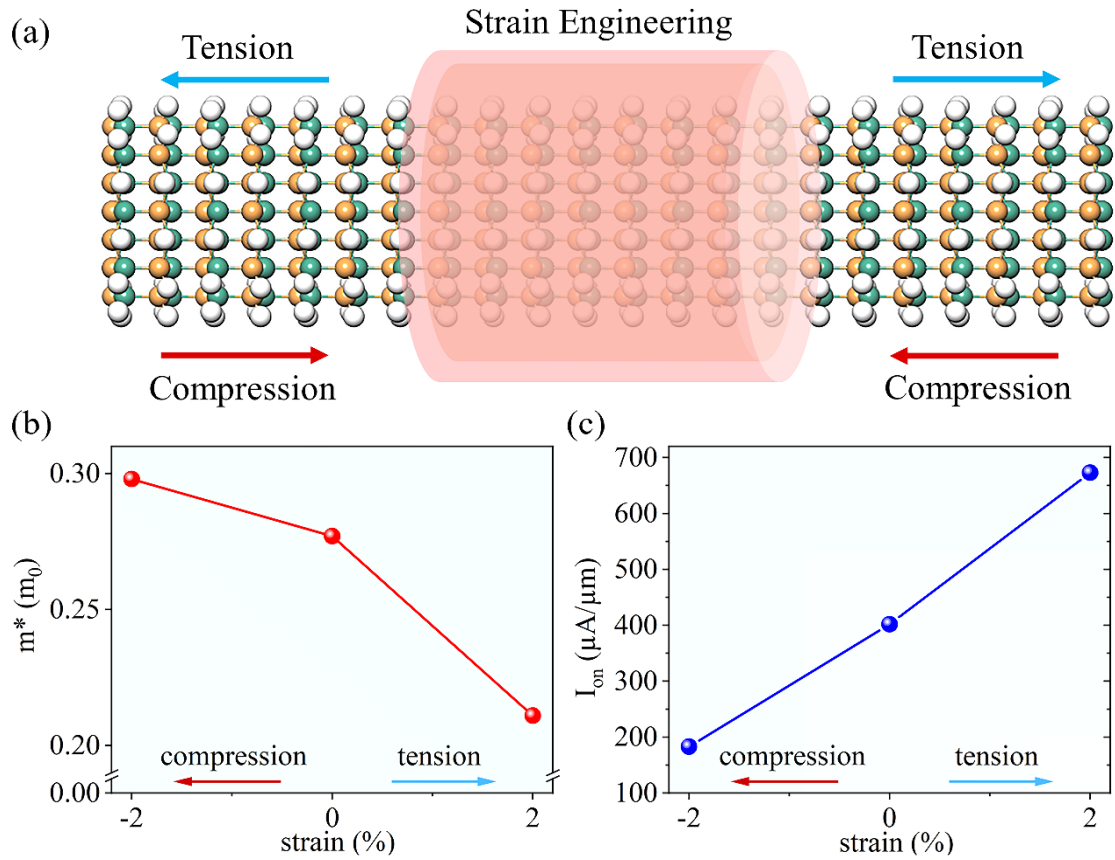

**Figure 6.** (a) Schematic diagram of the uniaxial strain on the GAA InP NW FET. (b) Effective mass *m\** of the InP NW versus uniaxial strain. The compressive strain is denoted by -2%, and the tensile strain is denoted by 2%. (c) $I_{on}$ of the 3-nm-$L_g$ GAA InP NW FETs against the uniaxial strain.

**TOC**

(a) GAA InP NW FET

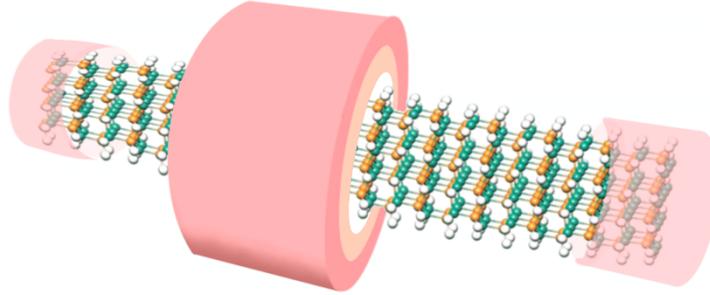

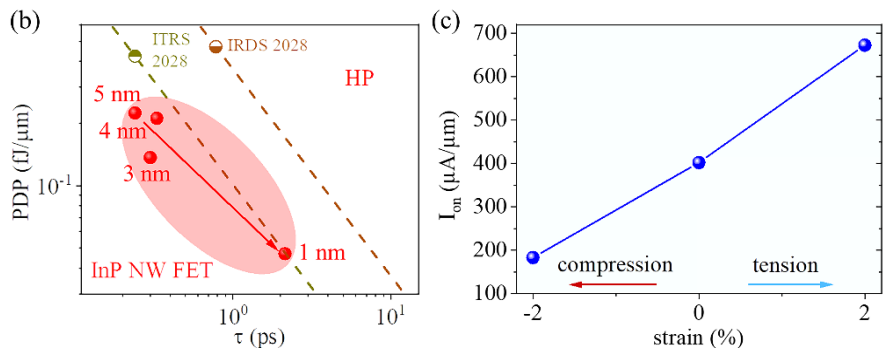

(b)

(c)